\documentclass{article}
\pdfoutput=1

\usepackage{authblk}
\usepackage[square,numbers]{natbib}
\usepackage{textcomp}

\usepackage{amsmath}
\usepackage{subfigure}
\newcommand{\comp}{\circ}
\usepackage{amssymb}
\usepackage[figuresright]{rotating}
\usepackage{color}

\usepackage[utf8]{inputenc} 
\usepackage[T1]{fontenc}    
\usepackage{hyperref}       
\usepackage{url}            
\usepackage{booktabs}       
\usepackage{amsfonts}       
\usepackage{nicefrac}       
\usepackage{microtype}      
\usepackage{xcolor}         
\usepackage{graphicx}       
\usepackage{pict2e}
\usepackage{xcolor}
\usepackage{wrapfig}
\usepackage{amssymb,amsmath,amsthm}
\theoremstyle{plain}

\theoremstyle{plain}

\theoremstyle{plain}

\theoremstyle{plain}

\providecommand{\assumptionname}{Assumption}
\providecommand{\lemmaname}{Lemma}
\providecommand{\propositionname}{Proposition}
\providecommand{\theoremname}{Theorem}

\usepackage{mathtools}

\usepackage{algorithm}
\usepackage[noend]{algpseudocode}

\usepackage{enumerate}

\usepackage[a4paper,left=3cm,right=3cm,top=3cm,bottom=3cm]{geometry}

\title{3D Tensor-based Deep Learning Models for Predicting Option Price}
\author{Muyang Ge}
\author{Shen Zhou}
\author{Shijun Luo}
\author{Boping Tian}
\affil{Department of Probability Statistics and Operation Control, \\School of Mathematics, Harbin Institute of Technology}
\date{}
\begin{document}

\maketitle

\begin{abstract} 
  Option pricing is a significant problem for option risk management and trading. In this article, we utilize a framework to present financial data from different sources. The data is processed and represented in a form of 2D tensors in three channels. Furthermore, we propose two deep learning models that can deal with 3D tensor data. Experiments performed on the Chinese market option dataset prove the practicability of the proposed strategies over commonly used ways, including B-S model and vector-based LSTM.
\end{abstract}

\section{Introduction}

In the area of option pricing, perhaps the most famous and widely used model is Black-Scholes (B-S) formula , which was raised in 1973. This formula is essentially the solution of a special heat conduction equation. B-S model makes several assumptions as a premise, including the Geometric Brownian Motion of the stock price and No Arbitrage Principle. However, in real option trading, these assumptions do not always hold, which leads to the bias of the prediction.

Meanwhile, the machine learning method establishes an algorithmic model in determining option price. The economic variables are utilized as input data, while the output is the option price. The complex relationship between input and output is obtained by learning a large amount of data by the algorithm itself, without considering the economic principle of this model. 

Deep learning (also known as deep structured learning) is part of a broader family of machine learning methods based on Artificial Neural Networks (ANN). Deep learning methods are more suitable for high-dimensional and complex data than traditional machine learning methods.

There have been many researches in the area of option pricing using machine learning \cite{culkin2017machine}, including deep learning strategies \cite{2018Research}. Hutchinson proposed a nonparametric method for asset pricing of financial derivatives by using machine learning, which is more accurate and effective. Maliaris and Salchenberger raised a model based on neural network, and the deviation of estimated option price is reduced compared with B-S model \cite{malliaris1993neural}. However, there is a disadvantage in using neural network to predict option price, that is, the over fitting of neural networks. Gradojevic et al. studied a nonparametric Fuzzy Neural Network (FNN) model to price European call options in the Standard \& Poor's 500 index. The experimental result shows that FNN always maintains excellent off sample pricing ability compared with other option pricing models. It shows that modularization improves the generalization ability of standard feedforward neural network \cite{gradojevic2009option}.

Financial data usually has the characteristics of time series. Recurrent Neural Networks (RNN) is a deep neural network specially designed for processing sequential data. Long Short-Term Memory (LSTM) is one of the most famous RNN architectures used in the field of deep learning. Therefore, many researches focused on estimating option price by employing LSTM and other RNN structures\cite{itkin2019deep, tan2019tensor, wei2020cnn}.

In this work, we start from processing and representing the option market data using a special data framework. Afterwards, two different models involving CNN, RNN, and Conv-LSTM are proposed innovatively to deal with the financial data in the form of 3D tensor. In the end, we conduct the empirical analysis using dataset from Chinese financial market to verify the practicability.

\section{Related Work}
\subsection{Tensor-based information framework}

This is a framework raised from the study of predicting stock price \cite{li2016tensor, sun2017predicting, atkins2018financial}. A common strategy in previous studies is to concatenate the features of multiple information sources (or modes) into one compound feature vector. This ignores the correlation of market information from different sources. To solve this problem, \cite{li2016tensor} proposed a framework that simulates the informational environment in respect of three types of information: firm-specific mode, event-specific mode and sentiment mode. In each mode of information, several measurements are selected to give a feature vector. For example, in firm-specific mode, five features have been used to capture the future business value of a firm: stock price, trading volume, turnover, price-to-earnings ratio and price-to-book ratio.

After the feature extraction, the investor information at time $t$ is represented with a 3D tensor $X_t\in \mathbb{R}^{I_1\times I_2\times I_3}$, where $I_1$, $I_2$ and $I_3$ are the number of features in three different modes of information.

However, this framework also has drawbacks. The most severe one is that the author organized the data tensor $X_t$ as follows:

\begin{itemize}
    \item $a_{i_1,1,1}$ , $1\leqslant i_1\leqslant I_1$ are the value of the fundamental features;
    \item $a_{2,i_2,2}$ , $1\leqslant i_2\leqslant I_2$ are the value of the event features;
    \item $a_{3,3,i_3}$ , $1\leqslant i_3\leqslant I_3$ are the value of the sentiment features;
    \item Other elements are set to zeros originally.
\end{itemize}

We can see from the above that most elements of this data tensor are set to zeros at first, which forces $X_t$ to be very sparse. When $X_t$ is input into the deep learning model, it is difficult for neural network to capture the useful information. Moreover, it wastes the computational resource to deal with a tensor full of zeros.

Even so, this model is effective in preserving the multifaceted and interrelated characteristics of market information. In this article, we explore the feasibility of applying this framework to the study in option pricing by making some modifications.

\subsection{Conv-LSTM}

Convolutional LSTM (Conv-LSTM) is a classical variation of traditional Fully-Connected LSTM (FC-LSTM, or LSTM) \cite{shi2015convolutional, tran2015learning}. It was raised to solve the problem in precipitation forecast, which predicts the precipitation in the following $K$ hours given previous $J$ observations:
\[
\tilde{X}_{t+1},\cdots ,\tilde{X}_{t+K}=\underset{X_{t+1},\cdots ,X_{t+K}}{\mathrm{arg}\max}p\left( X_{t+1},\cdots ,X_{t+K}|\hat{X}_{t-J+1},\hat{X}_{t-J+2},\cdots ,\hat{X}_t \right) 
\]

Obviously, we need to consider it from the perspective of time series and spatial relationship. The contribution of Conv-LSTM is to combine the convolution operation, which can extract spatial features, with the LSTM network, which can extract temporal features.

In traditional FC-LSTM, the input, hidden state and the memory cell are all 1D vectors. The operations of a single LSTM cell are shown below:

\[
i_t=\sigma \left( W_{xi}X_t+W_{hi}H_{t-1}+W_{ci}\comp C_{t-1}+b_i \right) 
\]
\[
f_t=\sigma \left( W_{xf}X_t+W_{hf}H_{t-1}+W_{cf}\comp C_{t-1}+b_f \right) 
\]
\[
\,\,C_t=f_t\comp C_{t-1}+i_t\comp \tanh \left( W_{xc}X_t+W_{hc}H_{t-1}+b_c \right) 
\]
\[
o_t=\sigma \left( W_{xo}X_t+W_{ho}H_{t-1}+W_{co}\comp C_t+b_o \right) 
\]
\[
H_t=o_t\comp \tanh \left( C_t \right) 
\]

In Conv-LSTM, however, we replace part of the matrix multiplication with convolution operation, in order to use it to reflect the spatial correlation inside the data:

\[
i_t=\sigma \left( W_{xi}*X_t+W_{hi}*H_{t-1}+W_{ci}\comp C_{t-1}+b_i \right) 
\]
\[
f_t=\sigma \left( W_{xf}*X_t+W_{hf}*H_{t-1}+W_{cf}\comp C_{t-1}+b_f \right) 
\]
\[
C_t=f_t\comp C_{t-1}+i_t\comp \tanh \left( W_{xc}*X_t+W_{hc}*H_{t-1}+b_c \right) 
\]
\[
o_t=\sigma \left( W_{xo}*X_t+W_{ho}*H_{t-1}+W_{co}\comp C_t+b_o \right) 
\]
\[
H_t=o_t\comp \tanh \left( C_t \right) 
\]
Where $*$ represents 2D convolution.

In this case, the input, hidden state and the memory cell are 2D tensors in several channels. Conv-LSTM considers both the time and spatial features, and many of the spatiotemporal forecasting problems take it as the building block. 

In this article, we use this model to capture the space characteristics of the financial data and achieve higher predictive accuracy. Convolution will be applied as a basic tool to find the features hidden in the 3D tensor data. However, we will replace the 2D convolution in this model by 1D convolution, since we lack one dimension of data to directly input the dataset into the original Conv-LSTM. 

\section{Model Structure}

There are varies kinds of information that can influence the option market such as fundamental data, data of price and data of Greeks. The classical strategies for processing data from different sources are to compromise them into a single vector and use the traditional method to make analysis. These methods fail to capture the correlation inside the different information of option market. Using a tensor-based representation of data effectively avoid the problems above.

Moreover, in this section, two deep learning models that can deal with 3D tensor data are built to catch the local feature and the long-term features hidden in the financial dataset.

\subsection{Feature extraction}

We utilize three different sources of data in this model: fundamental data, data of price and data of Greeks. From each of the information sources we select \textit{five} variables and thus there are totally \textit{fifteen} variables. Specifically,

\subsubsection*{Fundamental data}

The fundamental data reflects the intrinsic value of an option. The parameters we select are \textbf{spot price}, \textbf{strike price}, \textbf{day to expire}, \textbf{call or pull} (option) and \textbf{implicit volatility}. Note that they are all variables used in B-S formula. Nevertheless, we omit risk free rate, which is also calculated in B-S.

\subsubsection*{Data of price}

Several variables about price will also offer some information to predict value of settle price. Here, we select four of them: \textbf{previous settle price}, \textbf{settle price change}, \textbf{theory price} and \textbf{theory margin}. To make the number of dimensions of data from different sources be identical, we add the variable \textbf{inventory} in this part.

\subsubsection*{Data of Greeks}

For a general pricing model, the price of an option is determined by the factors listed in \emph{fundamental data} section. In these variables, except the strike price is constant, the change of any other factors will cause the change of corresponding option value, and also bring investment risk to the option. The Greeks are the quantities reflecting the sensitivity of the price of derivatives such as options to a change in underlying parameters on which the value of derivatives is dependent. In this work, the following five Greeks are used:

\noindent \textbf{Delta}: variation of option price caused by variation of spot price.

\noindent \textbf{Gamma}: variation of Delta caused by variation of spot price.

\noindent \textbf{Theta}: variation of option price caused by passing of time.

\noindent \textbf{Vega}: variation of option price caused by variation of implicit volatility.

\noindent \textbf{Rho}: variation of option price caused by variation of risk free rate.

\subsection{Feature representation}
The input data at the time $t$ is a 3D tensor with three channels of 2D tensors (matrices). The three channels represent the three different information sources: fundamental data, data of price and data of Greeks. In each channel, the data is organized as illustrated in Figure 1. In these matrices, the row $O_i$ represents the \emph{ith} option, and the column $F_{j}$ stands for the \emph{jth} input feature of this option \cite{wei2020cnn}. 

The target data, or in other words label, is option price (settle price). We choose to predict the price in each day using the previous 10 days' data. So input data of time $t-9$ to $t$ is matched with the target at time $t$.

We can conclude from above that for a single day, the shape of the input is $\left( C, N, D \right)$, where $C$ = 3, $D$ = 5 represent number of channels and number of variables respectively, and $N$ is the number of observations. For the whole dataset of input, the shape is $\left( T,C,N,D \right)$, where $T$ denotes the timeline. The corresponding target data has a shape $\left( N,1 \right)$.

\begin{figure}[htpb]
	\centering    
	{
		\begin{minipage}{1\textwidth}
            \centering
   			\includegraphics[width=10cm]{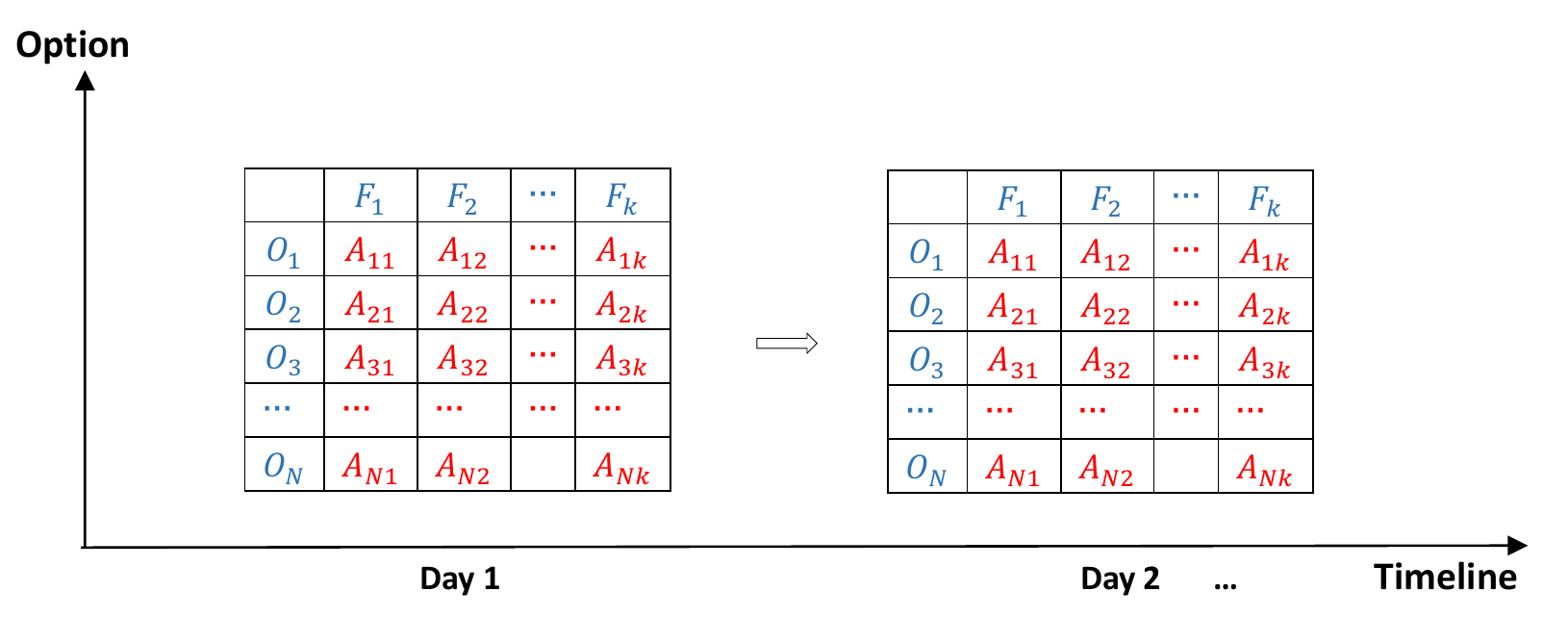}
		\end{minipage}
    }
	{
		\begin{minipage}{1\textwidth}
            \centering
			\includegraphics[width=6cm]{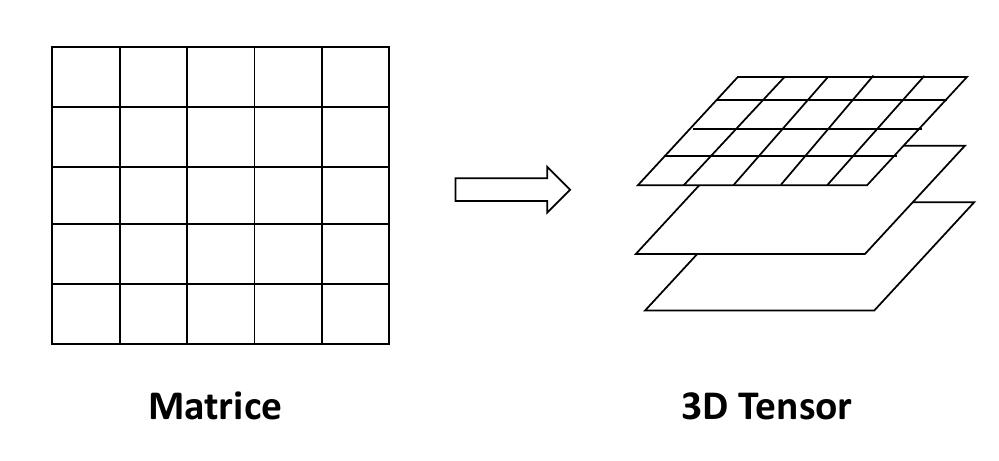} 
		\end{minipage}
	}	
	\caption{Feature representation \cite{wei2020cnn}} 
	\label{fig:1}  
\end{figure}

\subsection{CNN+RNN model}
This is a model combining CNN and RNN together, which has been widely used in spatiotemporal prediction problems \cite{alfarraj2019semi, alfarraj2019semisupervised}. The proposed model in the workflow consists of three main submodules, namely spatial modeling, temporal modeling and regression.

\subsubsection*{Spatial modeling}
The \emph{spatial modeling} submodule consists of a series of 1D convolutional blocks with different dilation. The output of each convolutional blocks is then concentrated and input into another convolutional block whose dilation is one. The 1D convolution here is along two directions of the data: timeline ($T$) and variables ($D$).

\subsubsection*{Temporal modeling}
The \emph{temporal modeling} submodule consists of three Gated Recurrent Units (GRU), which is a kind of RNN model, similar to LSTM model. Each of them is of one layer and bidirectional. The hidden channels of the three GRUs are 8, 16, and 16, respectively. For this submodule, we change the data shape into $\left( T, N, C\times D \right) $ to use GRU. 

\subsubsection*{Regression}
The last submodule is \emph{regression}. It contains a GRU unit and a linear mapping layer (fully connected layer). The main role of this module is to change the output shape of CNN and RNN into the shape of target: $\left( N,1 \right) $.

\begin{figure}[htpb]\label{ArchCR}
\centering
\includegraphics[width = 1\textwidth]{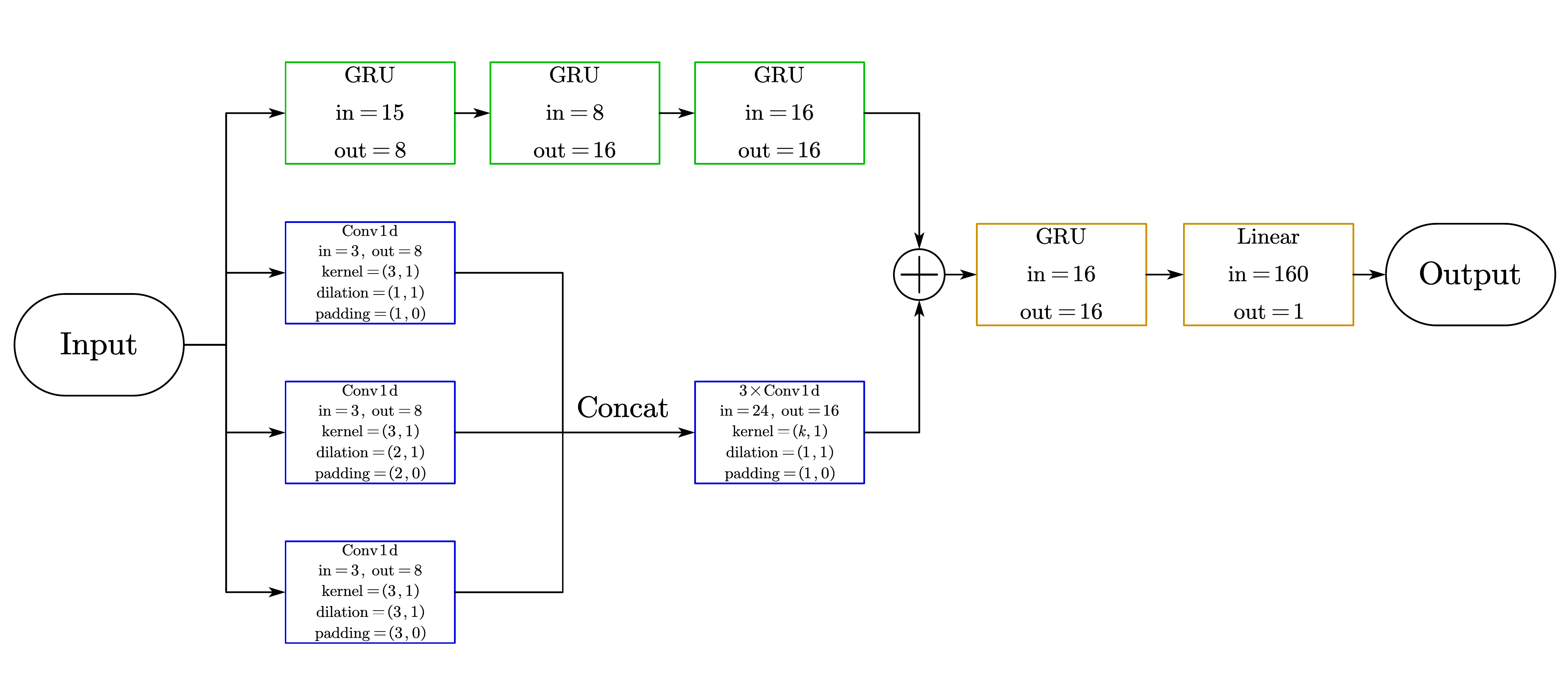}
\caption{The architecture of the CNN+RNN model}
\end{figure}

\subsection{Conv-LSTM model}

In the original paper \cite{shi2015convolutional}, Conv-LSTM is created to deal with the spatiotemporal sequence-forecasting problem. The form of data here is also $X _t\in \mathbb{R}^{C\times D\times E}$ ($E$ is symmetric to $D$), which should be viewed as $C$ channels of 2-dimension data (the shape is $D\times E$). Hence, the shape of the whole dataset for input is $\left( T, C, N, D, E \right) $. In each layer and each cell, the 2D convolution operates along the dimensions $D$ and $E$.

However, the shape of option dataset lacks one dimension needed in Conv-LSTM. It only has a shape of $\left( T, C, N, D \right) $. To solve this, we naturally think of replacing the 2D convolution in Conv-LSTM with 1D convolution. The 1D convolution is along the component $D$. Hence, the corresponding inputs $X_1\cdots \cdots X_t$, cell outputs $C_1\cdots \cdots C_t$, hidden outputs $H_1\cdots \cdots H_t$, and gates $i_t, f_t, o_t$ are all dealt with 1D convolution. 
The 1D convolution is more useful to gather the correlation of market information, and hence performs better than just using vector-based inputs.

The concrete computational process is as follows:

\[
i_t=\sigma \left( W_{xi}*X_t+W_{hi}*H_{t-1}+W_{ci}\comp C_{t-1}+b_i \right) 
\]
\[	
f_t=\sigma \left( W_{xf}*X_t+W_{hf}*H_{t-1}+W_{cf}\comp C_{t-1}+b_f \right) 
\]
\[
\,\,       C_t=f_t\comp C_{t-1}+i_t\comp \tanh \left( W_{xc}*X_t+W_{hc}*H_{t-1}+b_c \right) 
\]
\[
o_t=\sigma \left( W_{xo}*X_t+W_{ho}*H_{t-1}+W_{co}\comp C_t+b_o \right) 
\]
\[
H_t=o_t\comp \tanh \left( C_t \right) 
\]

Where $*$ denotes the 1D convolution between data (hidden state) and convolution kernel. 

As a comparison, we also reshape our input data as $\left( N, 1, T, C\times D \right) \,\,=\,\,\left( N,1,10,15 \right) $. That is, concentrating all variables into one channel. This is to test our strategy of feature representation.

\begin{figure}[htpb]\label{ArchConvLSTM}
\centering
\includegraphics[width = 0.8\textwidth]{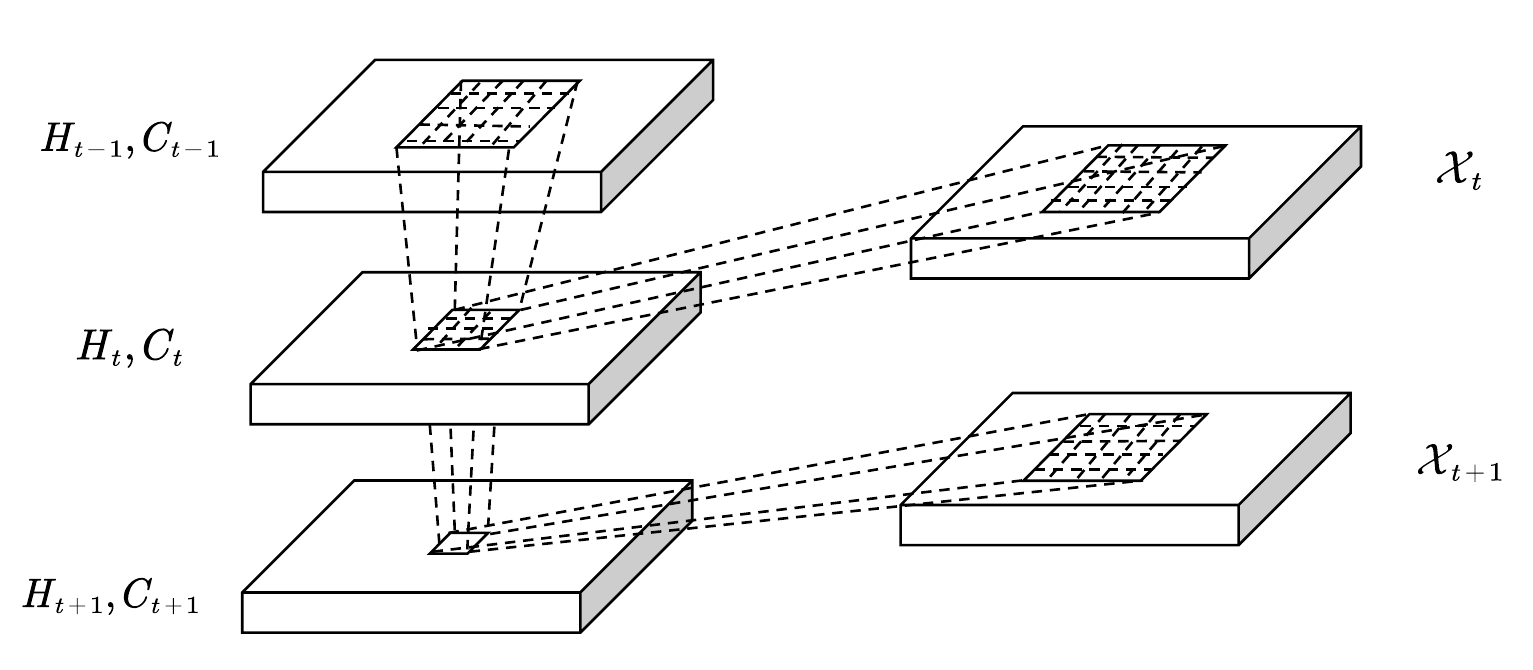}
\caption{The architecture of the Conv-LSTM model \cite{shi2015convolutional}}
\end{figure}

\section{Experiment}
\subsection{Experimental data}
In our experiments, 50ETF option and stock option data collected from Chinese stock and fund market from April 2018 to June 2020 are used. The original dataset totally includes 55047 observations and 34 variables. Those options with an effective period of at least 20 days are selected. After that, there are 829 options left. We utilize the first 600 options as training data and the other 229 options as test data. As stated in section 3.2, this dataset is transformed into correct shape. In the end, training data with 33210 observations and test data with 11386 observations are obtained. Data in this experiment is acquired from Yahoo Finance\footnote{\url{https://finance.yahoo.com}}.

In case the neural network tends to be imbalanced and dissymmetric, each variable in the dataset needs to be normalized before the experiments. To be specific:
\[
\tilde{x}\,\,=\,\,\left( x-\mu \right) /\sigma , 
\]
\[
\mu =\frac{1}{n}\sum_{i=1}^N{x_i},\ \sigma ^2=\frac{1}{n}\sum_{i=1}^N{\left( x_i-\mu \right) ^2}
\]

\subsection{Empirical analysis}

Under the environment of Pytorch, we built the two deep learning models raised before. For comparison, we also constructed the traditional B-S model and LSTM network using the variables mentioned. The loss function employed is the mean squared error (MSE) and the learning method is Adam. The mini-batch size and the learning rate are set to 64 and 0.0001. CNN+RNN model is trained for 100 epochs, while Conv-LSTM, LSTM are both trained for 200 epochs.

The prediction results are shown below. Only the first 120 days of option price in the test set is plotted on the pictures.

\begin{figure}[htpb]\label{BS}
\centering
\includegraphics[width = 0.8\textwidth]{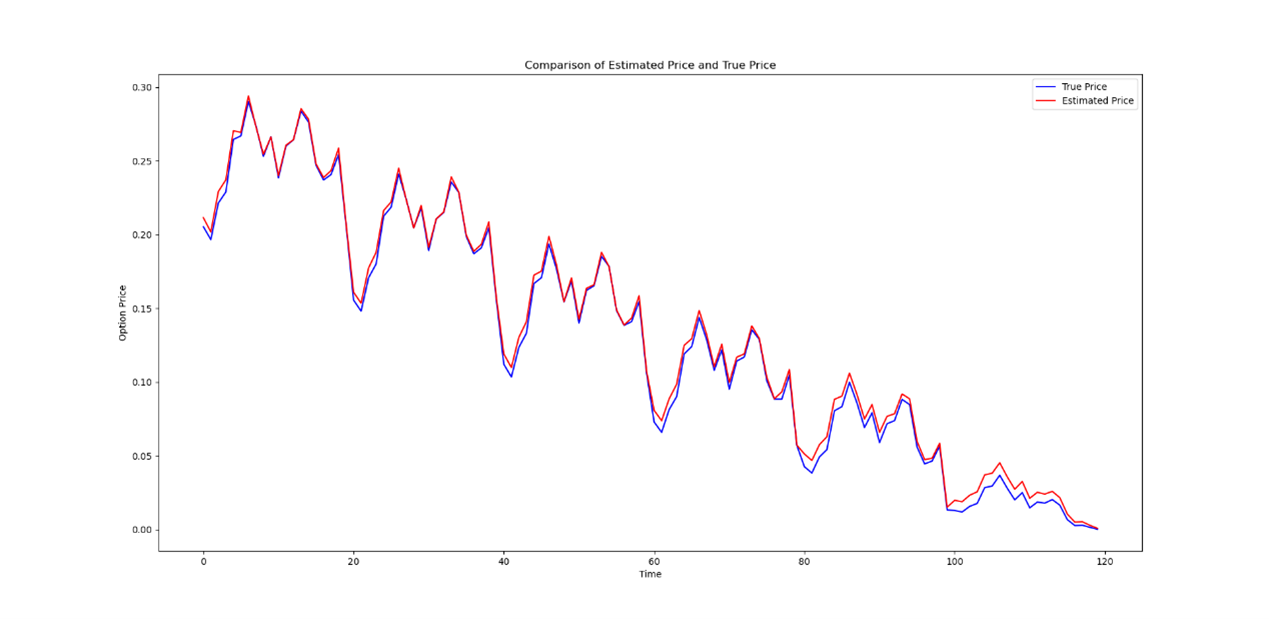}
\caption{Prediction result of B-S model}
\end{figure}

\begin{figure}[htpb]\label{LSTM}
\centering
\includegraphics[width = 0.8\textwidth]{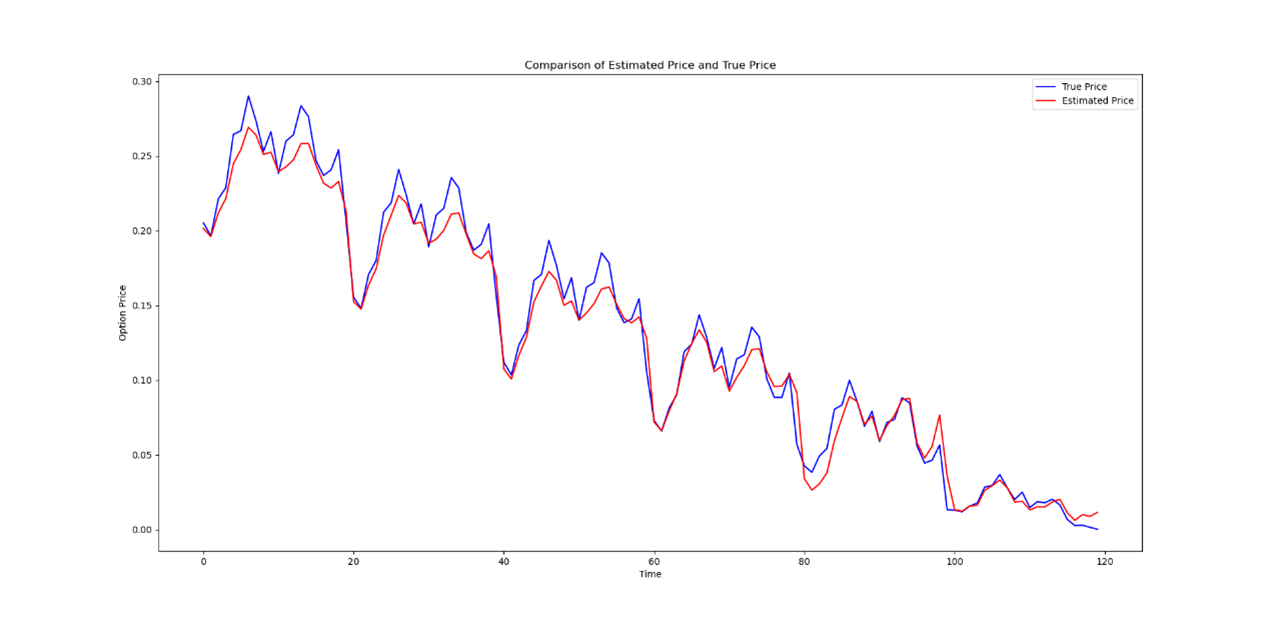}
\caption{Prediction result of LSTM model}
\end{figure}
	
\begin{figure}[htpb]\label{CRNN}
\centering
\includegraphics[width = 0.8\textwidth]{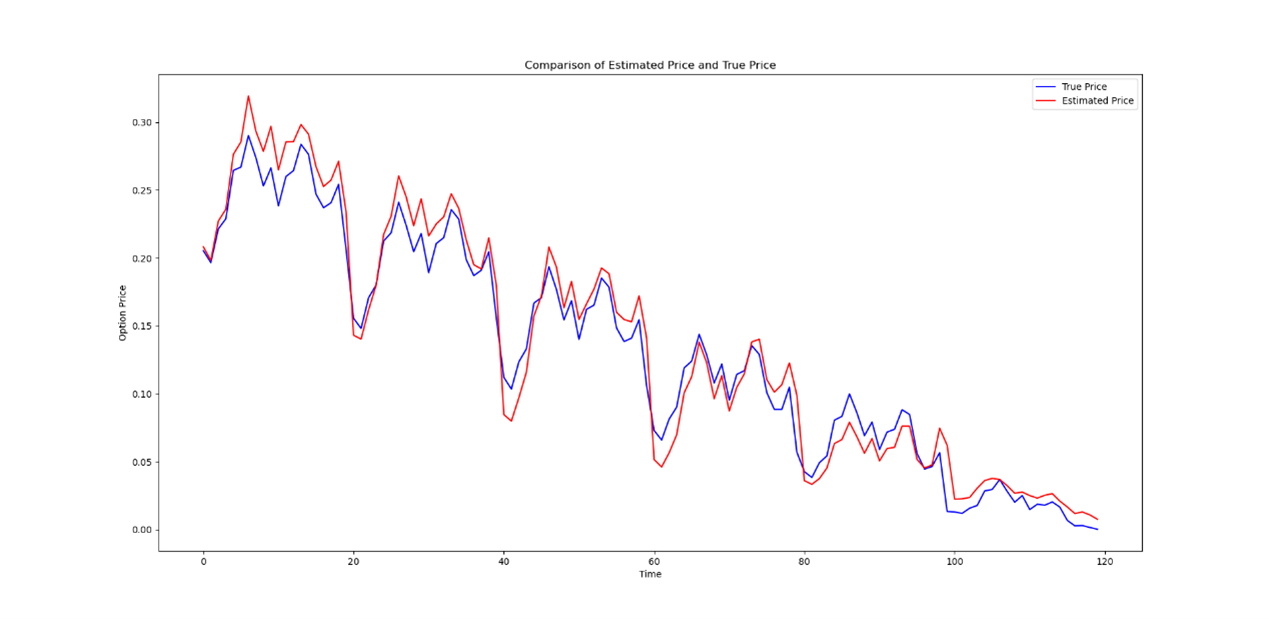}
\caption{Prediction result of CNN+RNN model}
\end{figure}

\begin{figure}[htpb]\label{ConvLSTM1}
\centering
\includegraphics[width = 0.8\textwidth]{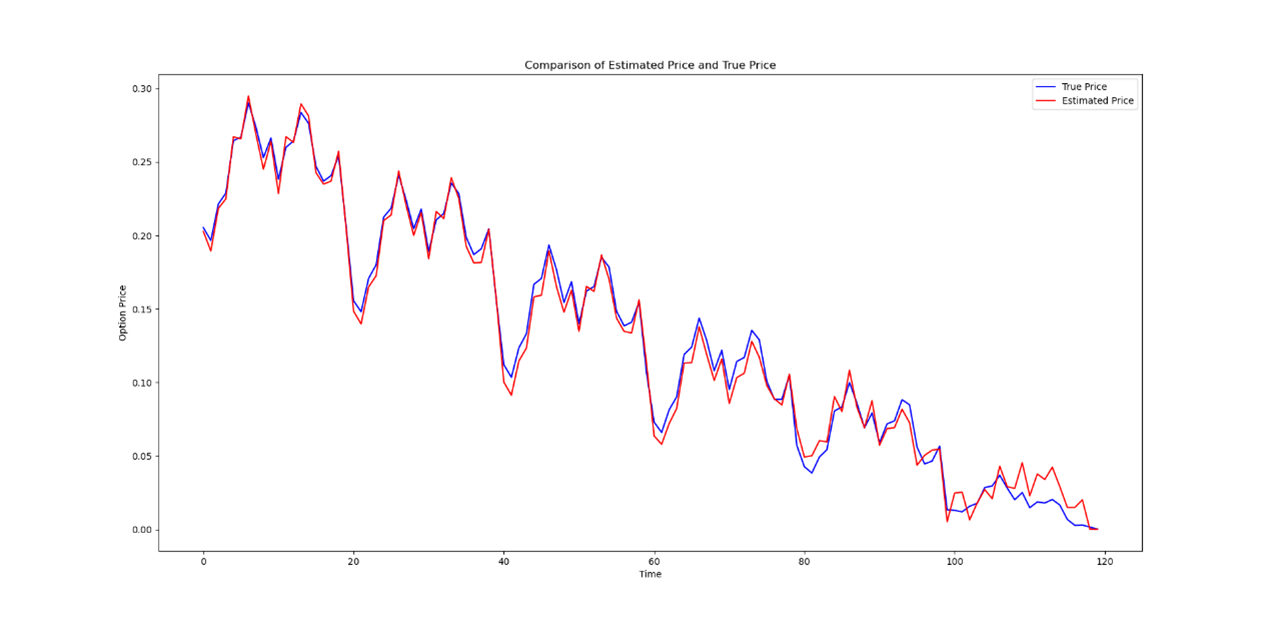}
\caption{Prediction result of Conv-LSTM model (with only 1 channel) }
\end{figure}

\begin{figure}[ht]\label{ConvLSTM3}
\centering
\includegraphics[width = 0.8\textwidth]{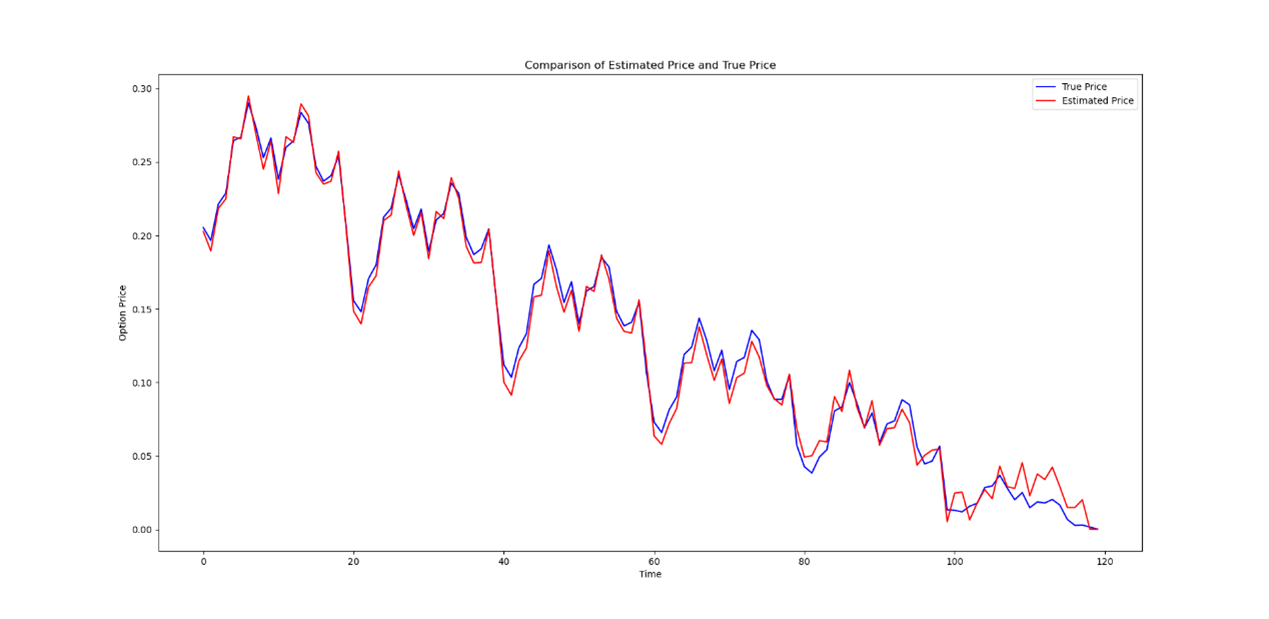}
\caption{Prediction result of Conv-LSTM model (with 3 channels)}
\end{figure}
	
It can be seen from the pictures that B-S performs not so good when it comes to the nadir of the price curve. For LSTM, the prediction is bad at the peak of the curve. CNN+RNN model seems to perform badly at both the peak and the nadir. As for Conv-LSTM, the prediction is not accurate when the true option price is very low.

To achieve a more detailed description of different models' predictive capability, the four different measurements are calculated: mean squared error (MSE), root mean square error (RMSE), mean absolute error (MAE), and mean absolute percentage error (MAPE). The calculation formulas are given below:
\[
	\begin{aligned}
		\mathrm{MSE}=\frac{1}{N}\sum_{i=1}^N{\left( y_i-\hat{y}_i \right) ^2}\qquad & \mathrm{RMSE}=\sqrt{\frac{1}{N}\sum_{i=1}^N{\left( y_i-\hat{y}_i \right) ^2}}\\
		\mathrm{MAP}=\frac{1}{N}\sum_{i=1}^N{\left| y_i-\hat{y}_i \right|}\,\,\phantom{^2}\qquad & \mathrm{MAPE}=\frac{1}{N}\sum_{i=1}^N{\left| \frac{y_i-\hat{y}_i}{y_i} \right|}
	\end{aligned}
\]

In addition, the Pearson correlation coefficient (PCC) between the true price and the estimated price is calculated. The results are listed in Table 1.

\begin{table}[h]
	\label{tab:RoM5}
	\centering
	\begin{tabular}[t]{cccccc}
		\toprule
		\  & MSE & RMSE & MAP & MAPE & PCC \\
		\midrule
		B-S & 0.00082 & 0.02872 & 0.02123 & 0.47957 & 0.99472 \\
		LSTM & 0.00027 & 0.01641 & 0.01165 & 2.06448 & 0.96584\\
		CNN-RNN & 0.00025 & 0.01585 & 0.01227 & 2.83609 & 0.95609\\
		Conv-LSTM (1C) & 0.00028 & 0.01659 & 0.01093 & 1.16076 & 0.96427\\
		Conv-LSTM (3C) & 0.00019 & 0.01396 & 0.00848 & 1.18657 & 0.96941\\
		\bottomrule
	\end{tabular}
	\caption{Result of models in five different measurements}
\end{table}

From the table, PCC of each model exceeds 0.95, which means the strong linear relationship between the true price and the estimated price.

Both the Conv-LSTM model and the CNN+RNN model surpass the LSTM model and B-S model in MSE and RMSE. Conv-LSTM model has the best performance in MSE, RMSE and MAP. Nevertheless, in MAPE, \emph{no} deep learning model shows better performance than the traditional B-S model, despite that their MAP is lower. It seems that B-S tends to give a value close to zero if the true price is very low, that decreases the MAPE of its prediction. This is a direction to work on when we construct the deep learning models for option pricing in the future.

The code of this article can be found on Github\footnote{\url{https://github.com/Tom-900/3D-Tensor-based-Deep-Learning-Models-for-Predicting-Option-Price.git}}.

\section{Conclusion}

Nowadays, deep learning is widely used in various fields, and it has become a hot topic in finance. In this article, we propose a different data framework composed of information from different sources and improve the LSTM neural network. After that, we compare them with LSTM model and B-S model in the empirical analysis. The result shows that the prediction ability of new models is better than that of traditional B-S model and LSTM model. 

Artificial intelligence will have great application value in the financial field. The algorithmic model based on deep learning method does not need to build a complex mathematical model, yet the prediction effect is more accurate. Deep learning in finance is reshaping the financial services industry significantly. Applying the state-of-art artificial intelligence methods in financial field will be of great importance in the future.

\section{Acknowledgement}
The authors declare that there is no conflict of interests regarding the publication of this article. This research is financed by NSFC grant 91646106.
  
\bibliographystyle{unsrt}
\bibliography{reference.bib}
\end{document}